\documentclass[twocolumn]{article}

\newcommand{\ket}[1]{|#1\rangle}
\newcommand{\bra}[1]{\langle#1|}
\newcommand{\Hil}[1]{{\cal H}_{#1}}
\newcommand{\Pauli}[1]{{\cal P}_{#1}}
\newcommand{\Cliff}[1]{{\cal C}_{#1}}
\newcommand{\rec}{{\cal R}}
\newcommand{\errors}{{\cal E}}
\newcommand{\wt}{{\rm wt}}  

\newtheorem{theorem}{Theorem}
\newtheorem{definition}{Definition}

\begin{document}

\title{Quantum Error Correction and Fault-Tolerance}
\author{Daniel Gottesman}
\date{}

\maketitle

\begin{abstract}
I give an overview of the basic concepts behind quantum error
correction and quantum fault tolerance.  This includes the quantum
error correction conditions, stabilizer codes, CSS codes,
transversal gates, fault-tolerant error correction, and the
threshold theorem.
\end{abstract}

\section{Quantum Error Correction}

Building a quantum computer or a quantum communications device in
the real world means having to deal with errors.  Any qubit stored
unprotected or one transmitted through a communications channel
will inevitably come out at least slightly changed. The theory of
quantum error-correcting codes has been developed to counteract
noise introduced in this way. By adding extra qubits and carefully
encoding the quantum state we wish to protect, a quantum system
can be insulated to great extent against errors.

To build a quantum computer, we face an even more daunting task:
If our quantum gates are imperfect, everything we do will add to
the error.  The theory of fault-tolerant quantum computation tells
us how to perform operations on states encoded in a quantum
error-correcting code without compromising the code's ability to
protect against errors.

In general, a quantum error-correcting code is a subspace of a
Hilbert space designed so that any of a set of possible errors can
be corrected by an appropriate quantum operation.  Specifically:
\begin{definition}
Let $\Hil{n}$ be a $2^n$-dimensional Hilbert space ($n$ qubits),
and let $C$ be a $K$-dimensional subspace of $\Hil{n}$.  Then $C$
is an $((n,K))$ (binary) quantum error-correcting code (QECC)
correcting the set of errors $\errors = \{E_a\}$ iff $\exists
\rec$ s.t.\ $\rec$ is a quantum operation and $(\rec \circ
E_a)(\ket{\psi}) = \ket{\psi}$ for all $E_a \in \errors$,
$\ket{\psi} \in C$.
\end{definition}

$\rec$ is called the {\em recovery} or {\em decoding} operation
and serves to actually perform the correction of the state. The
decoder is sometimes also taken to map $\Hil{n}$ into an unencoded
Hilbert space $\Hil{\log K}$ isomorphic to $C$.  This should be
distinguished from the {\em encoding} operation which maps
$\Hil{\log K}$ into $\Hil{n}$, determining the imbedding of $C$.
The computational complexity of the encoder is frequently a great
deal lower than that of the decoder.  In particular, the task of
determining what error has occurred can be computationally
difficult (NP-hard, in fact), and designing codes with efficient
decoding algorithms is an important task in quantum error
correction, as in classical error correction.

This article will cover only binary quantum codes, built with
qubits as registers, but all of the techniques discussed here can
be generalized to higher-dimensional registers, or {\em qudits}.

To determine whether a given subspace is able to correct a given
set of errors, we can apply the quantum error-correction
conditions~\cite{BDSW,KL}:
\begin{theorem}
A QECC $C$ corrects the set of errors $\errors$ iff
\begin{equation}
\bra{\psi_i} E_a^\dagger E_b \ket{\psi_j} = C_{ab} \delta_{ij},
\end{equation}
where $E_a, E_b \in \errors$ and $\{\ket{\psi_i}\}$ form an
orthonormal basis for $C$.
\end{theorem}

The salient point in these error-correction conditions is that the
matrix element $C_{ab}$ does not depend on the encoded basis
states $i$ and $j$, which roughly speaking indicates that neither
the environment nor the decoding operation learns any information
about the encoded state. We can imagine the various possible
errors taking the subspace $C$ into other subspaces of $\Hil{n}$,
and we want those subspaces to be isomorphic to $C$, and to be
distinguishable from each other by an appropriate measurement. For
instance, if $C_{ab} = \delta_{ab}$, then the various erroneous
subspaces are orthogonal to each other.

Because of the linearity of quantum mechanics, we can always take
the set of errors $\errors$ to be a linear space: If a QECC
corrects $E_a$ and $E_b$, it will also correct $\alpha E_a + \beta
E_b$ using the same recovery operation.  In addition, if we write
any superoperator $\cal{S}$ in terms of its operator-sum
representation ${\cal S}(\rho) \mapsto \sum A_k \rho A_k^\dagger$,
a QECC that corrects the set of errors $\{A_k\}$ automatically
corrects ${\cal S}$ as well.  Thus, it is sufficient in general to
check that the error-correction conditions hold for a basis of
errors.

Frequently, we are interested in codes that correct any error
affecting $t$ or fewer physical qubits.  In that case, let us
consider tensor products of the Pauli matrices
\begin{eqnarray}
& I = \pmatrix{1 & 0 \cr 0 & 1},\ & X = \pmatrix{0 & 1 \cr 1 & 0}, \nonumber \\
& Y = \pmatrix{0 & -i \cr i & 0},\ & Z = \pmatrix{1 & 0 \cr 0 &
-1}.
\end{eqnarray}
Define the Pauli group $\Pauli{n}$ as the group consisting of
tensor products of $I$, $X$, $Y$, and $Z$ on $n$ qubits, with an
overall phase of $\pm 1$ or $\pm i$.  The {\em weight} $\wt(P)$ of
a Pauli operator $P \in \Pauli{n}$ is the number of qubits on
which it acts as $X$, $Y$, or $Z$ (i.e., not as the identity).
Then the Pauli operators of weight $t$ or less form a basis for
the set of all errors acting on $t$ or fewer qubits, so a QECC
which corrects these Pauli operators corrects all errors acting on
up to $t$ qubits.  If we have a channel which causes errors
independently with probability $O(\epsilon)$ on each qubit in the
QECC, then the code will allow us to decode a correct state except
with probability $O(\epsilon^{t+1})$, which is the probability of
having more than $t$ errors.  We get a similar result in the case
where the noise is a general quantum operation on each qubit which
differs from the identity by something of size $O(\epsilon)$.

\begin{definition}
The distance $d$ of an $((n,K))$ QECC is the smallest weight of a
nontrivial Pauli operator $E \in \Pauli{n}$ s.t.\ the equation
\begin{equation}
\bra{\psi_i} E \ket{\psi_j} = C(E) \delta_{ij}
\end{equation}
fails.
\end{definition}
We use the notation $((n,K,d))$ to refer to an $((n,K))$ QECC with
distance $d$.  Note that for $P, Q \in \Pauli{n}$, $\wt(PQ) \leq
\wt(P) + \wt(Q)$.  Then by comparing the definition of distance
with the quantum error-correction conditions, we immediately see
that a QECC corrects $t$ general errors iff its distance $d > 2t$.
If we are instead interested in {\em erasure} errors, when the
location of the error is known but not its precise nature, a
distance $d$ code corrects $d-1$ erasure errors.  If we only wish
to {\em detect} errors, a distance $d$ code can detect errors on
up to $d-1$ qubits.

One of the central problems in the theory of quantum error
correction is to find codes which maximize the ratios $(\log K)/n$
and $d/n$, so they can encode as many qubits as possible and
correct as many errors as possible.  Conversely, we are also
interested in the problem of setting upper bounds on achievable
values of $(\log K)/n$ and $d/n$. The quantum Singleton bound (or
Knill-Laflamme bound~\cite{KL}) states that any $((n,K,d))$ QECC
must satisfy
\begin{equation}
n-\log K \geq 2d-2.
\end{equation}
We can set a lower bound on the existence of QECCs using the
quantum Gilbert-Varshamov bound, which states that, for large $n$,
an $((n,2^k,d))$ QECC exists provided that
\begin{equation}
k/n \leq 1 - (d/n) \log 3 - h(d/n),
\end{equation}
where $h(x) = -x \log x - (1-x) \log (1-x)$ is the binary Hamming
entropy.  Note that the Gilbert-Varshamov bound simply states that
codes at least this good exist; it does not suggest that better
codes cannot exist.

\section{Stabilizer Codes}

In order to better manipulate and discover quantum
error-correcting codes, it is helpful to have a more detailed
mathematical structure to work with.  The most widely-used
structure gives a class of codes known as {\em stabilizer
codes}~\cite{CRSS,Gottesman}. They are less general than arbitrary
quantum codes, but have a number of useful properties that make
them easier to work with than the general QECC.

\begin{definition}
Let $S \subset \Pauli{n}$ be an Abelian subgroup of the Pauli
group that does not contain $-1$ or $\pm i$, and let $C(S) =
\{\ket{\psi}\ {\rm s.t.}\ P\ket{\psi} = \ket{\psi}\ \forall P \in
S \}$.  Then $C(S)$ is a stabilizer code and $S$ is its {\em
stabilizer}.
\end{definition}

Because of the simple structure of the Pauli group, any Abelian
subgroup has order $2^{n-k}$ for some $k$ and can easily be
specified by giving a set of $n-k$ commuting generators.

The codewords of the QECC are by definition in the $+1$-eigenspace
of all elements of the stabilizer, but an error $E$ acting on a
codeword will move the state into the $-1$-eigenspace of any
stabilizer element $M$ which anticommutes with $E$:
\begin{equation}
M \left( E \ket{\psi} \right) = - EM \ket{\psi} = - E \ket{\psi}.
\end{equation}
Thus, measuring the eigenvalues of the generators of $S$ tells us
information about the error that has occurred.  The set of such
eigenvalues can be represented as an $(n-k)$-dimensional binary
vector known as the {\em error syndrome}. Note that the error
syndrome does not tell us anything about the encoded state, only
about the error that has occurred.

\begin{theorem}
Let $S$ be a stabilizer with $n-k$ generators, and let $S^\perp =
\{E \in \Pauli{n}\ {\rm s.t.}\ [E,M] = 0\ \forall M \in S \}$.
Then $S$ encodes $k$ qubits and has distance $d$, where $d$ is the
smallest weight of an operator in $S^\perp \setminus S$.
\end{theorem}
We use the notation $[[n,k,d]]$ to a refer to such a stabilizer
code.  Note that the square brackets specify that the code is a
stabilizer code, and that the middle term $k$ refers to the number
of encoded qubits, and not the dimension $2^k$ of the encoded
subspace, as for the general QECC (whose dimension might not be a
power of $2$).

$S^\perp$ is the set of Pauli operators that commute with all
elements of the stabilizer.  They would therefore appear to be
those errors which cannot be detected by the code.  However, the
theorem specifies the distance of the code by considering $S^\perp
\setminus S$. A Pauli operator $P \in S$ cannot be detected by the
code, but there is in fact no need to detect it, since all
codewords remain fixed under $P$, making it equivalent to the
identity operation. A distance $d$ stabilizer code which has
nontrivial $P \in S$ with $\wt(P) < d$ is called {\em degenerate},
whereas one which does not is {\em non-degenerate}. The phenomenon
of degeneracy has no analogue for classical error correcting
codes, and makes the study of quantum codes substantially more
difficult than the study of classical error correction.  For
instance, a standard bound on classical error correction is the
Hamming bound (or sphere-packing bound), but the analogous quantum
Hamming bound
\begin{equation}
k/n \leq 1 - (t/n) \log 3 - h(t/n)
\end{equation}
for $[[n,k,2t+1]]$ codes (when $n$ is large) is only known to
apply to non-degenerate quantum codes (though in fact we do not
know of any degenerate QECCs that violate the quantum Hamming
bound).

An example of a stabilizer code is the $5$-qubit code, a
$[[5,1,3]]$ code whose stabilizer can be generated by
\begin{eqnarray}
& X \otimes Z \otimes Z \otimes X \otimes I, \nonumber \\
& I \otimes X \otimes Z \otimes Z \otimes X, \nonumber \\
& X \otimes I \otimes X \otimes Z \otimes Z, \nonumber \\
& Z \otimes X \otimes I \otimes X \otimes Z. \nonumber
\end{eqnarray}
The $5$-qubit code is a non-degenerate code, and is the smallest
possible QECC which corrects 1 error (as one can see from the
quantum Singleton bound).

It is frequently useful to consider other representations of
stabilizer codes.  For instance, $P \in \Pauli{n}$ can be
represented by a pair of $n$-bit binary vectors $(p_X|p_Z)$ where
$p_X$ is $1$ for any location where $P$ has an $X$ or $Y$ tensor
factor and is $0$ elsewhere, and $p_Z$ is $1$ for any location
where $P$ has a $Y$ or $Z$ tensor factor. Two Pauli operators $P =
(p_X|p_Z)$ and $Q = (q_X|q_Z)$ commute iff $p_X \cdot q_Z + p_Z
\cdot q_X = 0$.  Then the stabilizer for a code becomes a pair of
$(n-k) \times n$ binary matrices, and most interesting properties
can be determined by an appropriate linear algebra exercise.
Another useful representation is to map the single-qubit Pauli
operators $I$, $X$, $Y$, $Z$ to the finite field GF(4), which sets
up a connection between stabilizer codes and a subset of classical
codes on 4-dimensional registers.

\section{CSS codes}

CSS codes are a very useful class of stabilizer codes invented by
Calderbank and Shor, and by Steane~\cite{CS,Steane}.  The
construction takes two binary classical linear codes and produces
a quantum code, and can therefore take advantage of much existing
knowledge from classical coding theory. In addition, CSS codes
have some very useful properties which make them excellent choices
for fault-tolerant quantum computation.

A classical $[n,k,d]$ linear code ($n$ physical bits, $k$ logical
bits, classical distance $d$) can be defined in terms of an $(n-k)
\times n$ binary {\em parity check} matrix $H$ --- every classical
codeword $v$ must satisfy $Hv = 0$.  Each row of the parity check
matrix can be converted into a Pauli operator by replacing each
$0$ with an $I$ operator and each $1$ with a $Z$ operator.  Then
the stabilizer code generated by these operators is precisely a
quantum version of the classical error-correcting code given by
$H$.  If the classical distance $d = 2t+1$, the quantum code can
correct $t$ bit flip ($X$) errors, just as could the classical
code.

If we want to make a QECC that can also correct phase ($Z$)
errors, we should choose {\em two} classical codes $C_1$ and
$C_2$, with parity check matrices $H_1$ and $H_2$.  Let $C_1$ be
an $[n,k_1,d_1]$ code and let $C_2$ be an $[n,k_2,d_2]$ code.  We
convert $H_1$ into stabilizer generators as above, replacing each
$0$ with $I$ and each $1$ with $Z$.  For $H_2$, we perform the
same procedure, but each $1$ is instead replaced by $X$. The code
will be able to correct bit flip ($X$) errors as if it had a
distance $d_1$ and to correct phase ($Z$) errors as if it had a
distance $d_2$.  Since these two operations are completely
separate, it can also correct $Y$ errors as both a bit flip and a
phase error.  Thus, the distance of the quantum code is at least
$\min(d_1,d_2)$, but might be higher because of the possibility of
degeneracy.

However, in order to have a stabilizer code at all, the generators
produced by the above procedure must commute.  Define the dual
$C^\perp$ of a classical code $C$ as the set of vectors $w$ s.t.\
$w \cdot v = 0$ for all $v \in C$.  Then the $Z$ generators from
$H_1$ will all commute with the $X$ generators from $H_2$ iff
$C_2^\perp \subseteq C_1$ (or equivalently, $C_1^\perp \subseteq
C_2$).  When this is true, $C_1$ and $C_2$ define an
$[[n,k_1+k_2-n,d]]$ stabilizer code, where $d \geq \min(d_1,d_2)$.

The smallest distance-$3$ CSS code is the $7$-qubit code, a
$[[7,1,3]]$ QECC created from the classical Hamming code
(consisting of all sums of classical strings $1111000$, $1100110$,
$1010101$, and $1111111$).  The encoded $\ket{\overline{0}}$ for
this code consists of the superposition of all even-weight
classical codewords and the encoded $\ket{\overline{1}}$ is the
superposition of all odd-weight classical codewords.  The
$7$-qubit code is much studied because its properties make it
particularly well-suited to fault-tolerant quantum computation.

\section{Fault-Tolerance}

Given a QECC, we can attempt to supplement it with protocols for
performing fault-tolerant operations. The basic design principle
of a fault-tolerant protocol is that an error in a single location
--- either a faulty gate or noise on a quiescent qubit --- should
not be able to alter more than a single qubit in each block of the
quantum error-correcting code. If this condition is satisfied, $t$
separate single-qubit or single-gate failures are required for a
distance $2t+1$ code to fail.

Particular caution is necessary, as computational gates can cause
errors to propagate from their original location onto qubits that
were previously correct.  In general, a gate coupling pairs of
qubits allows errors to spread in both directions across the
coupling.

The solution is to use transversal gates whenever
possible~\cite{Shor}. A transversal operation is one in which the
$i$th qubit in each block of a QECC interacts only with the $i$th
qubit of other blocks of the code or of special ancilla states. An
operation consisting only of single-qubit gates is automatically
transversal.  A transversal operation has the virtue that an error
occurring on the $3$rd qubit in a block, say, can only ever
propagate to the $3$rd qubit of other blocks of the code, no
matter what other sequence of gates we perform before a complete
error-correction procedure.

In the case of certain codes, such as the $7$-qubit code, a number
of different gates can be performed transversally.  Unfortunately,
it does not appear to be possible to perform universal quantum
computations using just transversal gates.  We therefore have to
resort to more complicated techniques.  First we create special
encoded ancilla states in a non-fault-tolerant way, but perform
some sort of check on them (in addition to error correction) to
make sure they are not too far off from the goal.  Then we
interact the ancilla with the encoded data qubits using gates from
our stock of transversal gates and perform a fault-tolerant
measurement.  Then we complete the operation with a further
transversal gate which depends on the outcome of the measurement.

\section{Fault-Tolerant Gates}

We will focus on stabilizer codes.  Universal fault-tolerance is
known to be possible for any stabilizer code, but in most cases
the more complicated type of construction is needed for all but a
few gates. The Pauli group $\Pauli{k}$, however, can be performed
transversally on any stabilizer code.  Indeed, the set $S^\perp
\setminus S$ of undetectable errors is a boon in this case, as it
allows us to perform these gates.  In particular, each coset
$S^\perp/S$ corresponds to a different logical Pauli operator
(with $S$ itself corresponding to the identity).  On a stabilizer
code, therefore, logical Pauli operations can be performed via a
transversal Pauli operation on the physical qubits.

Stabilizer codes have a special relationship to a finite subgroup
$\Cliff{n}$ of the unitary group $U(2^n)$ frequently called the
{\em Clifford group}.  The Clifford group on $n$ qubits is defined
as the set of unitary operations which conjugate the Pauli group
$\Pauli{n}$ into itself; $\Cliff{n}$ can be generated by the
Hadamard transform, the CNOT, and the single-qubit $\pi/4$ phase
rotation ${\rm diag}(1,i)$. The set of stabilizer codes is exactly
the set of codes which can be created by a Clifford group encoder
circuit using $\ket{0}$ ancilla states.

Some stabilizer codes have interesting symmetries under the action
of certain Clifford group elements, and these symmetries result in
transversal gate operations.  A particularly useful fact is that a
transversal CNOT gate (i.e., CNOT acting between the $i$th qubit
of one block of the QECC and the $i$th qubit of a second block for
all $i$) acts as a logical CNOT gate on the encoded qubits for any
CSS code.  Furthermore, for the $7$-qubit code, transversal
Hadamard performs a logical Hadamard, and the transversal $\pi/4$
rotation performs a logical $-\pi/4$ rotation.  Thus, for the
$7$-qubit code, the full logical Clifford group is accessible via
transversal operations.

Unfortunately, the Clifford group by itself does not have much
computational power: it can be efficiently simulated on a
classical computer.  We need to add some additional gate outside
the Clifford group to allow universal quantum computation; a
single gate will suffice, such as the single-qubit $\pi/8$ phase
rotation ${\rm diag}(1,\exp(i\pi/4))$. Note that this gives us a
finite generating set of gates. However, by taking appropriate
products, we get an infinite set of gates, one that is dense in
the unitary group $U(2^n)$, allowing universal quantum
computation.

The following circuit performs a $\pi/8$ rotation, given an
ancilla state $\ket{\psi_{\pi/8}} = \ket{0} + \exp(i\pi/4)
\ket{1}$:

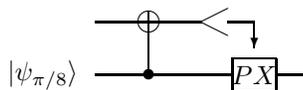
\begin{center}

\begin{picture}(120,60)

\put(0,14){\makebox(40,12){$\ket{\psi_{\pi/8}}$}}

\put(40,20){\line(1,0){52}} %
\put(40,40){\line(1,0){40}} %

\put(60,20){\circle*{4}} %
\put(60,20){\line(0,1){24}} %
\put(60,40){\circle{8}} %

\put(80,40){\line(2,1){10}} %
\put(80,40){\line(2,-1){10}} %

\put(90,40){\line(1,0){10}} %
\put(100,40){\vector(0,-1){10}} %

\put(92,14){\framebox(16,12){$PX$}} %
\put(108,20){\line(1,0){14}} %

\end{picture}

\end{center}

Here $P$ is the $\pi/4$ phase rotation ${\rm diag}(1,i)$, and $X$
is the bit flip.  The product is in the Clifford group, and is
only performed if the measurement outcome is $1$. Therefore, given
the ability to perform fault-tolerant Clifford group operations,
fault-tolerant measurements, and to prepare the encoded
$\ket{\psi_{\pi/8}}$ state, we have universal fault-tolerant
quantum computation.  A slight generalization of the
fault-tolerant measurement procedure below can be used to
fault-tolerantly verify the $\ket{\psi_{\pi/8}}$ state, which is a
$+1$ eigenstate of $PX$. Using this or another verification
procedure, we can check a non-fault-tolerant construction.

\section{Fault-Tolerant Measurement and Error Correction}

Since all our gates are unreliable, including those used to
correct errors, we will need some sort of fault-tolerant quantum
error correction procedure.  A number of different techniques have
been developed.  All of them share some basic features: they
involve creation and verification of specialized ancilla states,
and use transversal gates which interact the data block with the
ancilla state.

The simplest method, due to Shor, is very general but also
requires the most overhead and is frequently the most susceptible
to noise.  Note that the following procedure can be used to
measure (non-fault-tolerantly) the eigenvalue of any (possibly
multi-qubit) Pauli operator $M$: Produce an ancilla qubit in the
state $\ket{+} = \ket{0} + \ket{1}$. Perform a controlled-$M$
operation from the ancilla to the state being measured.  In the
case where $M$ is a multi-qubit Pauli operator, this can be broken
down into a sequence of controlled-$X$, controlled-$Y$, and
controlled-$Z$ operations.  Then measure the ancilla in the basis
of  $\ket{+}$ and $\ket{-} = \ket{0} - \ket{1}$.  If the state is
a $+1$ eigenvector of $M$, the ancilla will be $\ket{+}$, and if
the state is a $-1$ eigenvector, the ancilla will be $\ket{-}$.

The advantage of this procedure is that it measures just $M$ and
nothing more.  The disadvantage is that it is not transversal, and
thus not fault-tolerant. Instead of the unencoded $\ket{+}$ state,
we must use a more complex ancilla state $\ket{00\ldots0} +
\ket{11\ldots1}$ known as a ``cat'' state. The cat state contains
as many qubits as the operator $M$ to be measured, and we perform
the controlled-$X$, -$Y$, or -$Z$ operations transversally from
the appropriate qubits of the cat state to the appropriate qubits
in the data block.  Since, assuming the cat state is correct, all
of its qubits are either $\ket{0}$ or $\ket{1}$, the procedure
either leaves the data state alone or performs $M$ on it
uniformly. A $+1$ eigenstate in the data therefore leaves us with
$\ket{00\ldots0} + \ket{11\ldots1}$ in the ancilla and a $-1$
eigenstate leaves us with $\ket{00\ldots0} - \ket{11\ldots1}$. In
either case, the final state still tells us nothing about the data
beyond the eigenvalue of $M$.  If we perform a Hadamard transform
and then measure each qubit in the ancilla, we get either a random
even weight string (for eigenvalue $+1$) or an odd weight string
(for eigenvalue $-1$).

The procedure is transversal, so an error on a single qubit in the
initial cat state or in a single gate during the interaction will
only produce one error in the data.  However, the initial
construction of the cat state is not fault-tolerant, so a single
gate error then could eventually produce two errors in the data
block. Therefore, we must be careful and use some sort of
technique to verify the cat state, for instance by checking if
random pairs of qubits are the same.  Also, note that a single
phase error in the cat state will cause the final measurement
outcome to be wrong (even and odd switch places), so we should
repeat the measurement procedure multiple times for greater
reliability.

We can then make a full fault-tolerant error correction procedure
by performing the above measurement technique for each generator
of the stabilizer.  Each measurement gives us one bit of the error
syndrome, which we then decipher classically to determine the
actual error.

More sophisticated techniques for fault-tolerant error correction
involve less interaction with the data but at the cost of more
complicated ancilla states.  A procedure due to Steane uses (for
CSS codes) one ancilla in a logical $\ket{\overline{0}}$ state of
the same code and one ancilla in a logical $\ket{\overline{0}} +
\ket{\overline{1}}$ state.  A procedure due to Knill (for any
stabilizer code) teleports the data qubit through an ancilla
consisting of two blocks of the QECC containing an encoded Bell
state $\ket{\overline{00}} + \ket{\overline{11}}$.  Because the
ancillas in Steane and Knill error correction are more complicated
than the cat state, it is especially important to verify the
ancillas before using them.

\section{The Threshold for \\Fault-Tolerance}

In an unencoded protocol, even one error can destroy the
computation, but a fully fault-tolerant protocol will give the
right answer unless multiple errors occur before they can be
corrected.  On the other hand, the fault-tolerant protocol is
larger, requiring more qubits and more time to do each operation,
and therefore providing more opportunities for errors. If errors
occur on the physical qubits independently at random with
probability $p$ per gate or timestep, the fault-tolerant protocol
has probability of logical error for a single logical gate or
timestep at most $C p^2$, where $C$ is a constant that depends on
the design of the fault-tolerant circuitry (assume the quantum
error-correcting code has distance $3$, as for the $7$-qubit
code).  When $p < p_t = 1/C$, the fault-tolerance helps,
decreasing the logical error rate. $p_t$ is the {\em threshold}
for fault-tolerant quantum computation. If the error rate is
higher than the threshold, the extra overhead means that errors
will occur faster than they can be reliably corrected, and we are
better off with an unencoded system.

To further lower the logical error rate, we turn to a family of
codes known as {\em concatenated codes}~\cite{AB,Kitaev,KLZ}.
Given a codeword of a particular $[[n,1]]$ QECC, we can take each
physical qubit and again encode it using the same code, producing
an $[[n^2,1]]$ QECC.  We could repeat this procedure to get an
$n^3$-qubit code, and so forth.  The fault-tolerant procedures
concatenate as well, and after $L$ levels of concatenation, the
effective logical error rate is $p_t (p/p_t)^{2^L}$ (for a base
code correcting $1$ error). Therefore, if $p$ is below the
threshold $p_t$, we can achieve an arbitrarily good error rate
$\epsilon$ per logical gate or timestep using only ${\rm
poly}(\log \epsilon)$ resources, which is excellent theoretical
scaling.

Unfortunately, the practical requirements for this result are not
nearly so good. The best rigorous proofs of the threshold to date
show that the threshold is at least $2 \times 10^{-5}$ (meaning
one error per $50,000$ operations). Optimized simulations of
fault-tolerant protocols suggest the true threshold may be as high
as $5\%$, but to tolerate this much error, existing protocols
require enormous overhead, perhaps increasing the number of gates
and qubits by a factor of a million or more for typical
computations.  For lower physical error rates, overhead
requirements are more modest, particularly if we only attempt to
optimize for calculations of a given size, but are still larger
than one would like.

Furthermore, these calculations make a number of assumptions about
the physical properties of the computer.  The errors are assumed
to be independent and uncorrelated between qubits except when a
gate connects them.  It is assumed that measurements and classical
computations can be performed quickly and reliably, and that
quantum gates can be performed between arbitrary pairs of qubits
in the computer, irrespective of their physical proximity.  Of
these, only the assumption of independent errors is at all
necessary, and that can be considerably relaxed to allow
short-range correlations and certain kinds of non-Markovian
environments.  However, the effects of relaxing these assumptions
on the threshold value and overhead requirements have not been
well-studied.


\begin{thebibliography}{99}

\bibitem {AB}D.~Aharonov and M.~Ben-Or, ``Fault-tolerant quantum
computation with constant error rate,'' quant-ph/9906129.

\bibitem {BDSW}C.~Bennett, D.~DiVincenzo, J.~Smolin, and W.~Wootters,
``Mixed state entanglement and quantum error correction,'' Phys.\
Rev.\ A \textbf{54} (1996), 3824--3851; quant-ph/9604024.

\bibitem {CRSS}A.~R.~Calderbank, E.~M.~Rains, P.~W.~Shor, and
N.~J.~A.~Sloane, ``Quantum error correction via codes over
$\mathrm{GF}(4)$,'' IEEE Trans.\ Inform.\ Theory \textbf{44}
(1998), 1369--1387; quant-ph/9605005.

\bibitem {CS}A.~R.~Calderbank and P.~W.~Shor, ``Good quantum
error-correcting codes exist,'' Phys.\ Rev.\ A \textbf{54} (1996),
1098--1105; quant-ph/9512032.

\bibitem {Gottesman}D.~Gottesman, ``Class of quantum error-correcting
codes saturating the quantum Hamming bound,'' Phys.\ Rev.\ A
\textbf{54} (1996), 1862--1868; quant-ph/9604038.

\bibitem {Kitaev}A.~Y.~Kitaev, ``Quantum error correction with
imperfect gates,'' Quantum Communication, Computing, and
Measurement (Proc.\ 3rd Int.\ Conf.\ of Quantum Communication and
Measurement) (Plenum Press, New York, 1997), p.~181--188.

\bibitem {KL}E.~Knill and R.~Laflamme, ``A theory of quantum
error-correcting codes,'' Phys.\ Rev.\ A \textbf{55} (1997),
900--911; quant-ph/9604034.

\bibitem {KLZ}E.~Knill, R.~Laflamme, and W.~H.~Zurek, ``Resilient
quantum computation,'' Science \textbf{279} (1998), 342--345.

\bibitem {Shor}P.~W.~Shor, ``Fault-tolerant quantum
computation,'' Proc.\ 35th Ann.\ Symp.\ on Fundamentals of
Computer Science (IEEE Press, Los Alamitos, 1996), pp.~56--65;
quant-ph/9605011.

\bibitem {Steane}A.~M.~Steane, ``Multiple particle interference and
quantum error correction,'' Proc.\ Roy.\ Soc.\ London A
\textbf{452} (1996), 2551--2577; quant-ph/9601029.

\end{thebibliography}
\end{document}